\documentclass[12pt]{article}
\usepackage{epsfig}
\usepackage{graphicx,epic,eepic,latexsym,amsmath,xcolor}

\usepackage{amsmath,amsthm,amsfonts}
\usepackage{amssymb,latexsym}

\newtheorem{theorem}{Theorem}

\allowdisplaybreaks

\title{Benford's Law and Distractors in Multiple Choice Exams}
\author{Fred M. Hoppe}
\date{}
\begin{document}
\maketitle

\begin{abstract}

Suppose that in a multiple choice examination the leading digit of the correct options follows Benford's Law, 
while the leading digit of the distractors is uniform. 
Consider a strategy for guessing at answers that selects the option with the lowest leading digit with ties broken at random.   
We provide an expression for the probability that this strategy selects the correct option and a slight generalization to selecting the option with the lowest $r$ significant digit string.

\end{abstract}

\section{Introduction}

Benford's Law  \cite{newcomb, benford} is a probability distribution on the integers $1, 2, \cdots, 9$ that has been show to 
provide an excellent theoretical fit to the frequency of the leading digit in many data sets \cite{nigrini, bergerHill}.  
Recently, an interesting article \cite{slepkov} examined whether knowledge of Benford's Law could provide an advantage over 
pure guessing in multiple choice exams.    It was posited that the correct answers would have a Benford-like leading digit 
frequency while the distractors (the incorrect choices) would be artificial, possibly following a uniform leading digit 
distribution.  Since Benford's distribution is decreasing and  concentrates 
more than 0.60 probability at the digits 1, 2, 3, students might improve their scores from pure guessing by selecting 
the answer with the smallest leading digit.

In order to test this premise, the authors of \cite{slepkov} first examined a number of test banks for physics and 
chemistry books and found that that the leading digit of the correct answers could be well-approximated by Benford's Law, 
providing yet another manifestation of the pervasiveness of this distribution.  Knowing that Benford's Law was applicable 
to the correct answers,  they simulated 5,000 mock multiple choice questions in which the correct option followed Benford's 
Law while the distractors followed a uniform distribution of leading digits.  For each question they selected the option 
with the lowest leading digits.  If multiple options had the same lowest leading  digit, then a choice was made from 
them at random.    They found in the simulation that the expected scores for 3, 4, and 5 option examinations were 
0.51, 0.41, and 0.33 respectively.

They then applied the strategy of selecting the answer with the lowest leading digit to {\em actual} testbanks where 
it was unknown how the distractors were determined.
Surprisingly, they found for the 4-option questions (representing the majority of questions), that this strategy had a 
success rate of only 24.6\%, which is consistent with what would obtain picking answers uniformly at random.  
 
In order to understand this negative result, they then examined the distractors. Surprisingly they found that the distractors,
too, had a Benford-like distribution, which had the effect of mitigating the potential advantage of applying Benford's Law 
as an exam-taking strategy when a student was uncertain of an answer.

In this note, a closed-form solution is provided giving the probability of a correct response when the 
leading digit strategy is applied to exams where the correct answers follow Benford's Law while the distractors are uniform.

\section{Result}

Let $N$ denote the number of distractors in a multiple choice exam in which there is only one correct option per question 
(so there are $N+1$ options per question) and  let $\text{Bin}(k,p)$ be a binomial random variable on $k$ trials, success probability $p.$  Let
$$
b_i = \log_{10}\left(\frac{i+1}{i}\right), \hspace{1em} i = 1, 2, \cdots, 9
$$
represent the probabilities of the leading digit in Benford's Law.
Consider a multiple choice examination in which the leading digit of the correct responses follows Benford's Law while each of 
the distractors, independently of the correct answer and each other, has a leading digit  uniformly selected from $1, 2, \cdots, 9.$
Suppose the strategy of selecting the answer with the lowest leading digit is followed (with ties being broken at random).  
Denote by
$P_N^B$ the probability that this strategy selects the correct answer in a question.

\begin{theorem}
\begin{equation}\label{BenfordProb}
\begin{split} 
P^B_N &=  \sum_{i=1}^9 b_i \left(\frac{10-i}{9}\right)^{N}  E \left[ \frac{1}  {  \rm{Bin}(   N, \frac{1}{10-i}  ) +1}       \right] \\
& = \frac{1}{9^N(N+1)}   \sum_{i=1}^9 b_i [(10-i)^{N+1} - (9-i)^{N+1}]
\end{split}
\end{equation}
where $E$ denotes expectation.
\end{theorem}

\section{Proof}

For a given question, let  $X$ denote the leading digit of the correct answer and let $Y_1, \cdots, Y_N$ denote the leading digits 
of each of  the $N$ distractors, where, for definiteness, the distractors are ordered as they appear in the choice of options.
The key to the argument is to recognize that $P^B_N$ is the probability that the correct answer has the smallest leading digit, 
accounting for multiple options with the same leading digit.

We take $N=2$ first and consider all possible cases where the correct answer has a leading significant digit of 1.
\begin{equation*}
\begin{split}
P(X=1, Y_1 \ge 2, Y_2 \ge 2) &= b_1 \left(\frac{8}{9}\right)^2\\
P(X=1, Y_1 = 1, Y_2 \ge 2, \text{select correct answer} ) &= b_1 \left(\frac{1}{9}\right) \left(\frac{8}{9}\right) \left(\frac{1}{2}\right)\\
P(X=1, Y_1 \ge 2, Y_2 = 1, \text{select correct answer} ) &= b_1 \left(\frac{8}{9}\right) \left(\frac{1}{9}\right) \left(\frac{1}{2}\right)\\
P(X=1, Y_1 =1, Y_2 = 1, \text{select correct answer} ) &= b_1 \left(\frac{1}{9}\right)^2 \left(\frac{1}{3}\right)
\end{split}
\end{equation*}

The sum of these four terms is the probability of selecting the correct answer and the correct answer having leading digit 1.
The ratios $\frac{1}{9}, \frac{8}{9}$
can be recognized as the success and failure probabilities, respectively, of a binomial random variable $W_1$ on $N=2$ trials with success probability   $\frac{1}{9}.$ 
 Multiplication of these by the fractions at the extreme right results in the expected value $E\left(\frac{1}{W_1+1}\right),$  and therefore the sum of the probabilities equals
$$
b_1 E\left(\frac{1}{W_1+1}\right).
$$
This takes care of the case $X = 1.$  

For $X = 2$ the parallel  four equations above become
\begin{equation*}
\begin{split}
P(X=2, Y_1 \ge 3, Y_2 \ge 3) &= b_2 \left(\frac{7}{9}\right)^2\\
P(X=2, Y_1 = 2, Y_2 \ge 3, \text{select correct answer} ) &= b_2 \left(\frac{1}{9}\right) \left(\frac{7}{9}\right) \left(\frac{1}{2}\right)\\\
P(X=2, Y_1 \ge 3, Y_2 = 2, \text{select correct answer} ) &= b_2 \left(\frac{7}{9}\right) \left(\frac{1}{9}\right) \left(\frac{1}{2}\right)\\\
P(X=2, Y_1 =2, Y_2 = 2, \text{select correct answer} ) &= b_2 \left(\frac{1}{9}\right)^2 \left(\frac{1}{3}\right)\\\
\end{split}
\end{equation*}

Because the fractions  $\frac{1}{9} + \frac{7}{9} \ne 1$ we cannot apply a binomial distribution argument directly.  But
if we replace the denominator 9 by 8 wherever it appears  then the new fractions $\frac{1}{8}  +\frac{7}{8}= 1$ so the 
argument for $X=1$ can be applied and we merely need to account for the change in denominator by multiplying by  $\left(\frac{8}{9}\right)^2$ because of  the two appearances of 9 in the denominators.  We then find that the sum simplifies to 
$\left(\frac{8}{9}\right)^2  E\left(\frac{1}{W_2+1}\right)$
where  $W_2$  is a binomial random variable  on $N$ trials with success probability   $\frac{1}{8}.$

The sum of the probabilities for these four cases corresponding to $X=2$ is thus
$$
b_2  \left(\frac{8}{9}\right)^{2}  E\left(\frac{1}{W_2+1}\right).
$$
Continuing this way for the remaining values $X = 3, \ldots, 9$ and summing, we obtain the first line of (\ref{BenfordProb}) for $N=2$ distractors.
An analogous argument is easily seen to apply for $N$ distractors in general.  

The second line of (\ref{BenfordProb}) follows from substituting the binomial probabilities and simplifying.

Note that in place of  the probabilities of leading digits $\{b_i\}$ in (\ref{BenfordProb}) 
one can use any probability distribution, in particular the uniform for which (\ref{BenfordProb}) should, and does,  yield
$\frac{1}{N+1}$ for $N$ distractors.  

Table 1 gives the results of (\ref{BenfordProb})  both $P_N^B$ for the Benford distribution of the correct answer and 
$P_N^U$ for a uniform distribution of the correct answer.
The second column confirms the simulation results carried out in \cite{slepkov}.  Differences are due to the limited number (5,000) of simulated mock multiple choice exams in \cite{slepkov}.

\begin{table} [h]
  \caption{Probabilities of Choosing the Correct Option}
\begin{center}
\begin{tabular}{r  |r| r }
$N$ & $P^B_N$ & $P_N^U$\\
\hline
1 &  0.6733    & 0.5000\\
2 &   0.5291   & 0.3333\\
3 &   0.4421   & 0.2500\\
4 &   0.3821    & 0.2000\\
5 &   0.3375   & 0.1667\\
\end{tabular}
\end{center}
\end{table}

\section{Using the First $r$ Digits}

Equation (\ref{BenfordProb}) assumes that ties are broken at random. 
There is an extension of Benford's Law for the leading $r$  significant digits.  For a number with at least $r$ significant digits, denote by
$d = (d_1, \cdots, d_r)$  the sequence comprising the first, second, $\cdots, r^{th}$ significant digits, respectively.
Then Benford's distribution assigns probability $b_d$ to the sequence $d$
where
\begin{equation}\label{bd}
b_d = \log_{10}\left(\frac{1+\sum_{j=1}^r 10^{r-j} d_j}{\sum_{j=1}^r 10^{r-j} d_j}\right).
\end{equation}
From (\ref{bd}) we can find that the distribution of  each significant digit also decreases from 1 to 9.  As a result, rather than break ties at random, we may break ties according to which second leading digit is smallest.  This is equivalent to picking the answer with the smallest initial two significant
 digits, considered as an integer between 10 and 99, inclusive.  If there are ties with two significant digits, we then break them by picking the answer with the smallest initial three significant digits and so on.

This leads us to consider a multiple choice examination in which each correct answer and distractor has at least $r$ significant digits.  Assume that
the first $r$ leading significant digits of the correct responses follow Benford's Law while each of the distractors, independently of the correct answer and each other, has the first $r$ significant digits uniformly distributed on the set of $r$-digit integers
 \{$i: 10^{r-1} \le i \le 10^{r}-1\}.$  Suppose the strategy of selecting the answer with the lowest $r$ leading digits is followed (with ties again being broken at random).  Can we do better than the previous strategy for $r=1?$  
 
 Let $P^{B,r}_N$ denote the probability that this strategy selects the correct answer for a question.   
 The argument leading to Theorem 1 extends easily.  For simplicity we present the result when $r=2.$
\begin{theorem}
\begin{equation}\label{Benford2}
\begin{split} 
P^{B,2}_N &=  \sum_{i=10}^{99} b_i \left(\frac{100-i}{90}\right)^{N}  E \left[ \frac{1}  {  \rm{Bin}(   N, \frac{1}{100-i}  ) +1}       \right] \\
& = \frac{1}{90^N(N+1)}   \sum_{i=10}^{99} b_i [(100-i)^{N+1} - (99-i)^{N+1}]
\end{split}
\end{equation}
\end{theorem}

It turns out (see Table 2) that there is a slight benefit for the strategy with $r=2$ over $r=1$, but little additional improvement beyond over randomly selecting  when there are ties. The reason would appear to be that the significant digits subsequent to the first one have a distribution approximating the uniform, 
which is the distribution of the distractors.

\begin{table} [h]
  \caption{Probabilities of Choosing the Correct Option for $k=2, 3$}
\begin{center}
\begin{tabular}{r  |r| r }
$N$ & $P^{B,2}_N$ & $P^{B,3}_N$\\
\hline
1 &   0.6768    & 0.6768\\
2 &   0.5348   & 0.5349\\
3 &   0.4495   & 0.4495\\
4 &   0.3908    & 0.3909\\
5 &   0.3474   & 0.3475\\
\end{tabular}
\end{center}
\end{table}

\subsection*{Acknowledgments}

Thanks to George Wesolowsky for drawing my attention to the interesting paper \cite{slepkov}.

\bigskip

\noindent \textit{Department of Mathematics and Statistics, McMaster University, Hamilton, ON L8S 4K1, Canada. hoppe@mcmaster.ca}.

\end{document}